\documentclass[onecolumn,notitlepage]{revtex4-1}

\usepackage{graphics} 
\usepackage{graphicx} 
\usepackage{color} 
\usepackage{amsmath}
\usepackage{amssymb}
\usepackage{amsfonts}
\usepackage{sidecap}

\newcommand{\qmin}{{q_{\text{min}}}}
\newcommand{\qmax}{{q_{\text{max}}}}

\begin{document} 

\title{Quasi-universality in mixed counterions systems}

\author{Martin Trulsson$^{1,3}$}
\author{Ladislav \v{S}amaj$^{2}$} 
\author{Emmanuel Trizac$^3$} 

\affiliation{
$^1$Theoretical Chemistry, Lund University, Sweden \\
$^2$Institute of Physics, Slovak Academy of Sciences, Bratislava, Slovakia \\ 
$^3$LPTMS, CNRS, Univ. Paris-Sud, Universit\'e Paris-Saclay, 91405 Orsay, France
}

\begin{abstract}
The screening of plate-plate interactions by counterions is an age-old problem. We revisit 
this classic question when counterions exhibit a distribution of charges. While it is expected
that the long-distance regime of interactions is universal, the behaviour of the inter-plate
pressure at smaller distances should {\it a priori} depend rather severely on the nature of the
ionic mixture screening the plate charges. We show that is not the case, and that for
comparable
Coulombic couplings, different systems exhibit a quasi-universal equation of state.
\end{abstract}
\date{\today}

\maketitle

Colloidal suspensions are made up of macromolecules in a more or less polar solvent. The presence of charged groups 
at the surface of the macromolecules  may lead to repulsive interactions, providing a mechanism 
to counteract ubiquitous van der Waals attractions, that would otherwise lead to aggregation,
and an instability of the suspension \cite{RuSC}. It is however possible to minimize and sometime cancel
these van der Waals forces, e.g. by refractive index matching \cite{Yeti07}. Integrating over
microscopic degrees of freedom (solvent, microions), one obtains the free energy for any given fixed configuration
of macromolecules, from which the equilibrium force felt by each macroion derives 
\cite{Bell00,Levi02}. The resulting so-called effective potential is the object of central interest 
in the present contribution. We focus on the simplest possible setting where two charged planar
macromolecules confine the solvent and charge-compensating counterions in a slab. More specifically, we are interested
in the effect of counterion polydispersity, in the sense that they may bear different charges. A central result 
reported is that of a certain universality of the effective plate-plate potential. To illustrate the
statement, we will consider rather extreme counterion charge distribution $n(q)$: while counterions
naturally bear a multiple of the elementary charge ($q=1$ for monovalent ions, $q=2$ for divalent ions, etc.),
and therefore have a $n(q)$ with discrete support, we will address cases where $n(q)$ is continuous,
to maximize the effect of non-monodispersity. 

It is known that the effective plate-plate potential (or equivalently in the present case, the pressure),
turns from all-distance repulsive under small Coulombic couplings, to a more complex dependence under larger
coupling, with repulsion at small distances, attraction in some intermediate range, and repulsion again at large 
distances. When all ions have the same elementary charge $-e$, quantification of coupling is operated by a 
dimensionless parameter $\Xi$, which is a measure of inverse temperature $\beta$ or, equivalently, of the 
plate surface charge $\sigma e$. In a solvent of permittivity $\epsilon$, we have 
$\Xi \propto \sigma \ell_B^2$ where $\ell_B=\beta e^2/\epsilon$ is the Bjerrum length.
As a rule of thumb, mean-field holds for $\Xi<1$ \cite{Netz01}.
Whereas the prototypical problem with a unique type of counterions (referred to as the monodisperse case with $q=1$)
is well documented \cite{Netz01,ChWe06,HaLu10,SaTr11}, we present here a combination 
of analytical and numerical results for polydisperse systems. Counterions have 
charge $-qe$ with a density distribution $n(q)$. We seek for several types of universal features:
a) within a given family of polydisperse distribution $n(q)$, a rather down-to-earth 
type amounts to finding the proper set of reduced variables, making, if possible, the
reduced pressure independent of the polydispersity; b) we are interested in changing the type
of polydispersity, considering distinct families of functions $n(q)$. Given the qualitatively distinct behavior 
at small and large couplings, it is certainly essential here to keep track of this coupling
in the description.

Yet, it should be clear from the outset that there is another and broader level of universality
in the large distance ($d$) behavior of the pressure. Indeed, irrespective of coupling parameter,
the two plates interact weakly at large $d$. There, one can expect counterion to effectively ``dress'' the plates,
modifying their effective charge (that it is not necessary to define in a more
rigorous manner), with in between the plates, a weak density of ions remaining,
only those with smallest valence $\qmin$. Thus, we expect the physics of
interactions to fall at large $d$ in the mean-field category, which has the remarkable property 
to be independent of the macroion charge \cite{EvWe}. This is a consequence of the 
$(\ell_B d^2)^{-1}$ dependence of pressure, which does not leave room for any dependence 
on $\sigma$, for dimensional reasons. We expect this result to be valid
as long as one can define a Bjerrum length ($\qmin^2 \ell_B$) associated to the population of
smallest valence, which means as long as $\qmin >0$ \cite{rque50}. 

As alluded to, we consider a system of two plates at distance $d$, both having uniform surface charge
$\sigma e$, and neutralized by counter-ions in the slab $-d/2 \leq x \leq d/2$. 
Denoting $\qmax$ the maximum valence, electroneutrality 
requires that
\begin{equation}
\int_{\qmin}^{\qmax} \, n(q) \ q \, dq \, =\, 2 \sigma \ ;
\end{equation}
$n(q)$ is the total (surface) density if ions having valence $q$.
We start by weakly coupled systems, where Poisson-Boltzmann mean-field treatment
should hold \cite{Levi02,Netz01}. Introducing the dimensionless electrostatic potential $\phi$, one has to
solve 
\begin{equation}
\frac{d^2 \phi}{dx^2} \, =\, 4 \pi \ell_B \, \int_\qmin^\qmax f(q)\, q \, e^{q\phi(x)} \, dq
\label{eq:pb}
\end{equation}
which appears in the form of an implicit formulation. Indeed, the volume density of species
$q$ can be written $f(q) \exp(q\phi)$, and the normalization function $f(q)$ is
{\it a priori} unknown. It relates to the chosen $n(q)$ through
\begin{equation}
n(q) \, =\, \int_{-d/2}^{d/2} f(q)\, e^{q\phi(x)} \, dx.
\label{eq:n-f}
\end{equation}
Therefore, starting from $n(q)$, $f(q)$ is not known before Eq.
\eqref{eq:pb} has been solved, itself requiring the knowledge of 
$f(q)$. Since the large $d$ asymptotics is constrained by the argument presented 
above, we focus on short $d$ features. It is for that purpose useful 
to choose $\phi(0)=0$, where symmetry and Gauss theorem further impose 
$\phi'(0) = 0$ and $\phi'(d/2) = 4\pi\ell_B \, \sigma$. 
Taylor expanding $\phi$, which is even, we have
\begin{eqnarray}
&&\phi(x) \, \sim \, a_1 x^2 + a_2 x^4 + \ldots \\
\hbox{with  }~ && a_1 = 2 \pi \ell_B \int q \, f(q)\, dq \\
&& a_2 \, = \,\frac{2}{3} \, (\pi \ell_B)^2 \, \left[
\int q\,f(q) dq \right] \left[ \int q^2\, f(q)\, dq
\right] .
\end{eqnarray}
where the last two lines are obtained by inserting the expansion into Eq. \eqref{eq:pb}.
Restricting to the quadratic term gives the dominant contribution 
to the pressure, that reads
\begin{eqnarray}
\beta P \,=\, \int f(q) \, dq ,
\end{eqnarray}
meaning that within the present mean-field description, it is given
by the mid-plane ionic density, a classic result stemming from the fact that 
the electric field vanishes at this point.
Making use of Eq. \eqref{eq:n-f} allows us to write 
\begin{equation}
\beta P \, =\, \frac{1}{d} \, \int n(q) \, dq  \, - \,  \frac{\pi \ell_B}{6} 
\left[\int q\, n(q) \, dq
\right]^2 .
\end{equation}
Remembering electroneutrality, and defining a (Gouy) length as 
$\mu = [2\pi\ell_B \sigma \langle q \rangle]^{-1}$
where  $ \langle q \rangle$ is the mean valence of the distribution
\cite{rque51}, we introduce the reduced separation 
$\widetilde d = d/\mu$ and obtain
\begin{equation}
\widetilde P \, \equiv \, \frac{\beta P}{2\pi\ell_B \sigma^2} \, =\, 
\frac{2}{\widetilde d} -\frac{1}{3},
\label{eq:100}
\end{equation}
which appears in a polydispersity independent form. This suggests
to use $\mu$ defined from $\langle q \rangle$ as the relevant length 
and consider $\widetilde P$. It is also instructive to push the
expansion one order higher in distance.
After some algebra, this yields 
\begin{equation}
\widetilde P   \, =\, 
\frac{2}{\widetilde d} -\frac{1}{3} \, +\, \frac{2}{45} \, \frac{\langle q^2\rangle}{\langle q^3\rangle} \, \widetilde d,
\end{equation}
where it is seen that the linear contribution in $\widetilde d$ is $n(q)$-dependent,
and in this sense non-universal. Thus, even within mean-field, we should not
expect full  universality. However, we shall show below that for practical purposes, the
equation of state $\widetilde P(\widetilde d)$ is remarkably insensitive to the details of the
polydispersity function $n(q)$.

Before presenting our numerical results, we show that the same rescaling as in 
\eqref{eq:100} leads to a universal small distance behavior under strong coupling
(i.e. beyond mean-field). There, the discrete nature of counter-ions is essential
and every such ion appears far from its neighbors, in the sense
that the corresponding distance is larger than the slab width $d$
\cite{Varenna}. As a consequence, the problem is reducible to a single particle 
picture, where all species adopt a flat ($x$-independent) profile with density
$n(q)/d$, given that the plates create a vanishing electric field in the slab 
$-d/2\leq x \leq d/2$. The contact theorem \cite{contact1,contact2,contact3} 
then immediately leads to the pressure, in the form
\begin{equation}
\beta P \, =\, -2 \pi \ell_B\, \sigma^2 + \frac{1}{d} \, \int n(q) \,dq,
\end{equation}
so that we have
\begin{equation}
\widetilde P \, =\, \frac{2}{\widetilde d} -1.
\label{eq:Psc}
\end{equation}
The proximity with Eq. \eqref{eq:100}, with the same dominant order, simply stems from
the dominant entropy cost of confinement \cite{rque52}. However, unlike
Eq. \eqref{eq:Psc}, Eq. \eqref{eq:100} is unable to predict like-charge attraction
(i.e $P<0$). Indeed, such a phenomenon would happen for $\widetilde d > 6$ from
\eqref{eq:100}, a distance range where our expansion no longer holds. The bottom-line
is that mean-field pressure is always repulsive \cite{Neu,Sader,PRE}.
On the other
hand, Eq. \eqref{eq:Psc} indicates attraction for $\widetilde d >2$, which is validated 
by numerical simulations.

\begin{figure}[t!]
\includegraphics[scale=0.8]{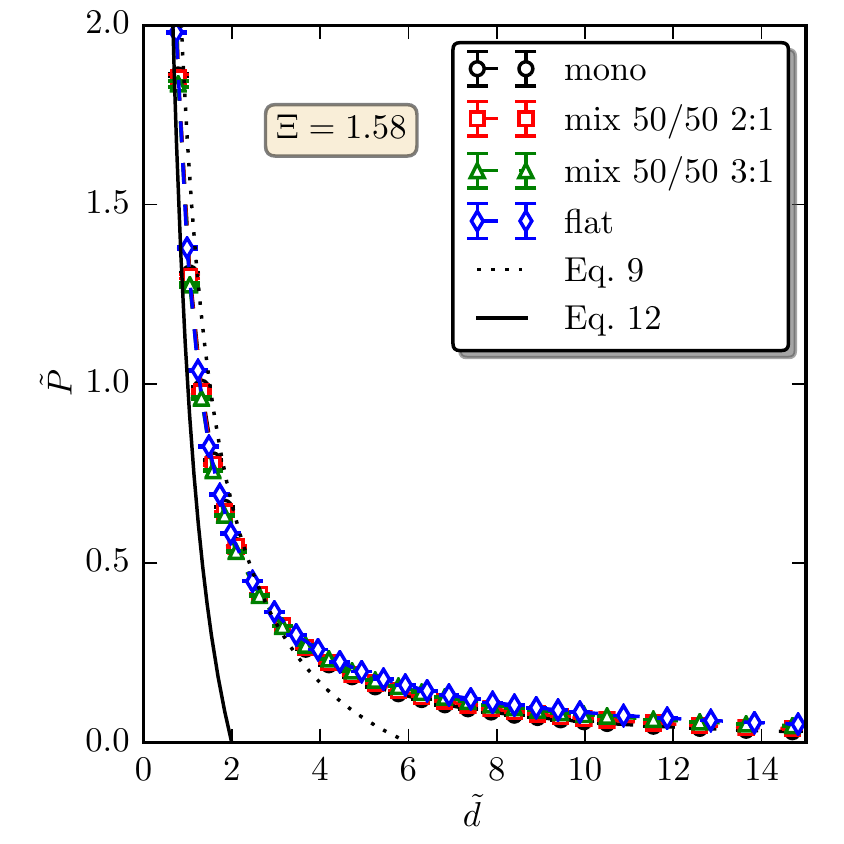}\\
\includegraphics[scale=0.8]{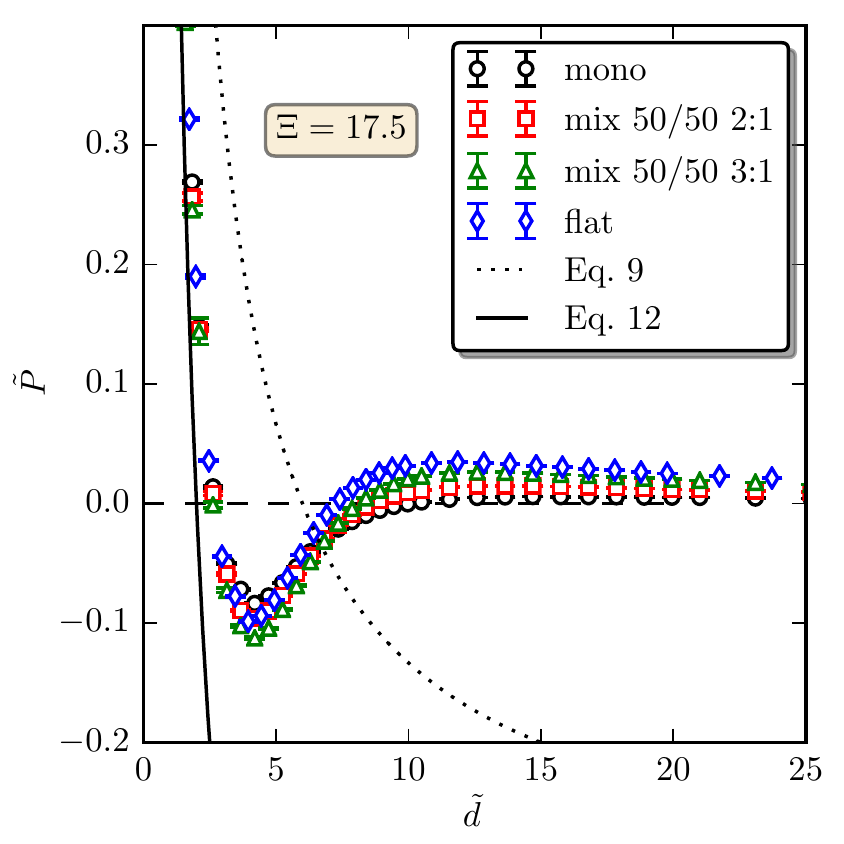}\\
\includegraphics[scale=0.8]{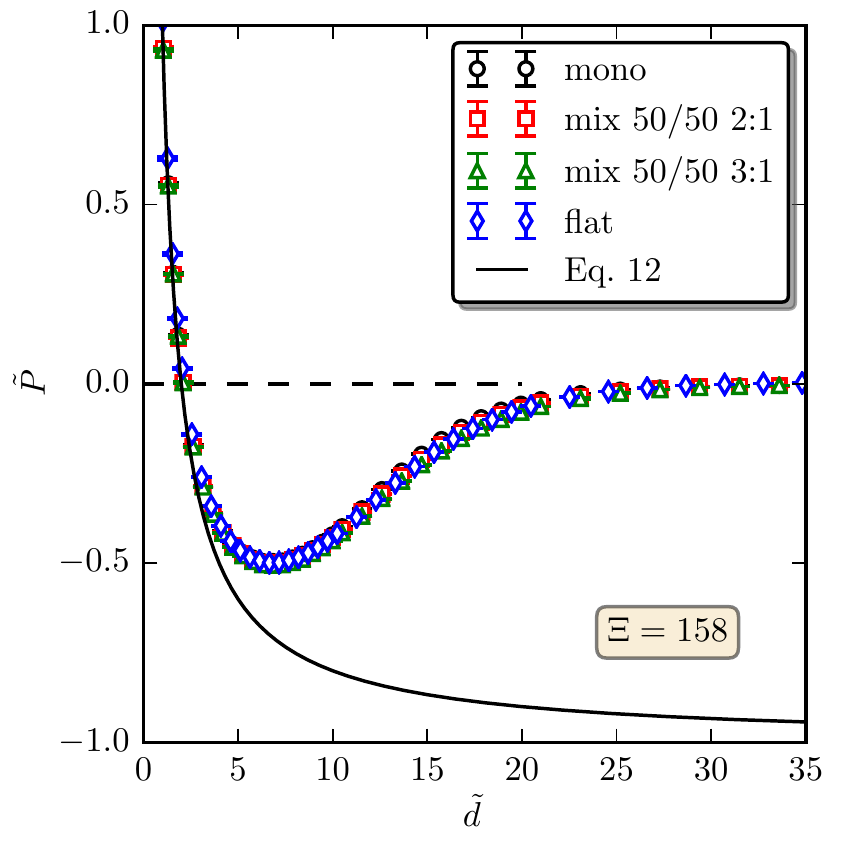}
\vspace{- 5 mm}
\caption{Normalized pressure versus normalized separation for three different coupling parameters $\Xi=1.58, 17.5$ and $158$. 
(Symbols) Monte-Carlo results (Lines) Analytic expressions.}
%\vspace{- 5 mm}
\label{Pressure}
\end{figure}

The above results indicate which dimensionless quantities should be computed when 
analyzing numerical simulation results. We have performed Monte-Carlo
simulations in a quasi-2D geometry. Long-ranged electrostatic interactions are handled with
Ewald summation techniques corrected for quasi-2D-dimensionality by introducing a vacuum
slab in the z-direction perpendicular to the surfaces \cite{Berkowitz,Mazars}. We verified that our vacuum slab is sufficiently wide not to influence the results. 
All simulations consisted of 512 point charges while the surfaces are modeled as structureless uniformly infinite plates,
in agreement with the analytical treatment.
Simulations were performed both for equimolar binary mixtures with charge ratios 2:1 and 3:1 (due to its relevance for realistic ionic systems) as well
as for a continuous flat distribution of charges, $q\in [0,q_{\rm max}]$.
Fig.~\ref{Pressure} shows the pressure curves for three different coupling parameters $\Xi= 2 \pi \langle q\rangle^3 \ell_B^2 \sigma$ which
all collapse onto each other at small plate-plate separations. For the larger coupling parameters ($\Xi=17.5$ and 158) one sees the well-documented although
counter-intuitive like-charge 
attraction, driven by ion-ion correlations. Impressively enough, even the attractive minimas are well-described, both in magnitude and width, 
for the same coupling parameter, irrespective of the mixture type. 
At $\Xi=17.5$, one does however observe small discrepancies, even at small separations, between the different mixtures of ions. 
These small discrepancies are more pronounced at intermediate and long separations, where the (modest) repulsive barrier seems
to be stronger (more repulsive) as the charge ratio of the binary mixture increases; it is magnified in the ``flat'' charge distribution.
Also, the other coupling parameters (1.58 and 158) exhibit minor discrepancies from intermediate to larger separations, but the variation induced
by ionic composition is small compared to the absolute values (note the different pressure scales in the figures). 
%First when the absolute values start to be small can one see a
%detectable difference both in relative terms between the mixtures. Maximum discrepancy might therefore be expected at the onset of the attractive interaction 
%occurring around $\Xi\simeq5$. 
Overall, the results show a decent agreement, and the proposed reduced parameter set (\textit{i.e.} $\Xi$, $\tilde{P}$, and $\tilde{d}$) 
captures well the dominant behavior. 
%The actual composition of the mixture seems to be sub-dominate for the plate-plate interaction.

%That this long-ranged repulsive interaction increases in strength and would penetrate into the short-ranged one and eventually blur the short-ranged 
%attraction is not unexpected as the charge ratio is increased. To understand this, one can consider the extreme case of 0:1 mixtures.
%especially from intermediate to large separations. 
 %reflecting that the long-ranged
%interaction might to be different compared to the short-ranged one and where a flatter distribution or a bigger charge ratio in the binary mixture 
%give a stronger repulsive interaction at intermediate to large separations. It is also for this intermediate coupling parameter where the minimas
%differ a bit, 

In conclusion, we have shown that the short-ranged interactions between two equally charged surfaces exhibit a robust quasi-universal equation of state,
irrespective of the type of mixture of counter-ions. This requires to introduce suitably defined reduced variables, where the inter-plate distance is
measured in units of a length, the inverse of which can be viewed as the mean inverse Gouy length ($1/\mu = 2 \pi \ell_B \,\sigma \langle q \rangle$),
where the average is taken over the counter-ion distribution $n(q)$. In addition, it is of course essential to keep track of Coulombic coupling.
While there is no clear-cut definition of a coupling parameter in the present polydisperse situation, we have shown that 
the parameter $\Xi = 2 \pi \langle q\rangle^3 \ell_B^2 \sigma$ allows for suitable rescaling. 
Our approach holds for a salt free system (counter-ions only) without 
any possibility of ion exchange. In an open system (e.g. in contact with a salt reservoir), 
the mixture composition would vary as a function of the plate separation. At short separations, 
co-ions would be expelled from the slab, to be released in the reservoir,
and one would be left with a salt-free system  favoring the high valency counter-ions.
In that limiting distance regime, our approach would be applicable.
Finally, we note that our results can be viewed as an (approximate) law
of corresponding states, where the mean valence of counter-ion present
plays a privileged role ($\langle q\rangle$). On the other hand, in the Deby-H\"uckel treatment
of electrolytic systems (where thus microscopic ions of both signs are present), the key quantity is the Debye length,
which involves a different moment, $\langle q^2\rangle$.

Acknowledgment. The support received

from VEGA Grant No. 2/0015/15 is acknowledged.

%Most of the work on strong-coupling and focusing on the
%short-ranged interactions could thereby be generalized also to any mixture of counter-ions.
%But to understand such systems it turns out that only the mean valency would be
%important to the system which changes essentially the average Gouy-Chapman length. 
%This sheds new light on the ion-ion correlation and strong coupling mechanism.


\begin{thebibliography}{10}
\bibitem{rque50}
When $\qmin \to 0$, we thus get a diverging pressure, which is indicative
of the fact that the pressure, in those cases, ceases to exhibit 
a power in distance with exponent $-2$, but can be longer range.
\bibitem{rque52}
Under strong coupling, computing the next order contribution in distance is a difficult task,
which requires knowledge of (at least) the local structure of the counterion system \cite{SaTr11}.
This is beyond our scope here.
\end{thebibliography}
\end{document}